\documentclass[twocolumn,prl,amsmath,amssymb,showpacs,superscriptaddress]{revtex4}
\usepackage{graphicx}
\usepackage{dcolumn}
\usepackage{bm}
\usepackage{latexsym}
\usepackage{amstext}
\usepackage{amsxtra}
\usepackage{color}

\newcommand{\bs}{\boldsymbol}

\begin{document}

\title{Mesoscopic ensembles of polar bosons in triple-well potentials}

\author{T.~Lahaye}
\affiliation{Universit\'e de Toulouse, UPS, Laboratoire Collisions Agr\'egats R\'eactivit\'e, IRSAMC ; F-31062 Toulouse, France}
\affiliation{CNRS, UMR 5589, F-31062 Toulouse, France}
\affiliation{5. Physikalisches Institut, Universit\"at Stuttgart, Pfaffenwaldring 57, 70550 Stuttgart, Germany}

\author{T.~Pfau}
\affiliation{5. Physikalisches Institut, Universit\"at Stuttgart, Pfaffenwaldring 57, 70550 Stuttgart, Germany}

\author{L. Santos}
\affiliation{Institut f\"ur Theoretische Physik, Leibniz Universit\"at Hannover, Appelstr. 2 D-30167, Hannover, Germany}

\date{April 27, 2010}

\begin{abstract}
Mesoscopic dipolar Bose gases in triple-well potentials offer a minimal system for the analysis of the non-local character of the dipolar interaction. We show that this non-local character may be clearly revealed by a variety of possible ground-state phases. In addition, an appropriate control of short-range and dipolar interactions may lead to novel scenarios for the dynamics of polar bosons in lattices, including the dynamical creation of mesoscopic quantum superpositions, which may be employed in the design of Heisenberg-limited atom interferometers.
\end{abstract}


\maketitle

Interparticle interactions are crucial in quantum gases~\cite{bloch2008}. They can usually be described by a short-range isotropic potential proportional to the scattering length $a$. Recently, \emph{dipolar} quantum gases, in which the \emph{long-range} and \emph{anisotropic} dipole-dipole interaction (DDI) between magnetic or electric dipole moments plays a significant or even dominant role, have attracted a lot of interest as they show fascinating novel properties~\cite{baranov2008,lahaye2009}. To date, dipolar effects have been observed experimentally only with atomic magnetic dipoles, being particularly relevant in Bose-Einstein condensates (BECs) of $^{52}$Cr where exciting new physics has been observed~\cite{lahaye2007,koch2008,lahaye2008,metz2009}. Dipolar effects have also been reported in spinor BECs~\cite{vengalattore2008}, and in $^{39}$K and $^7$Li BECs with $a=0$~\cite{fattori2008,Pollack2009}. Recent experiments with polar molecules~\cite{ni2008,Deiglmayr2008} open fascinating perspectives towards the realization of highly-dipolar gases.

Although a very clear and direct demonstration of the anisotropy of the DDI was given by the $d$-wave collapse of a Cr-BEC~\cite{lahaye2008,metz2009}, an equivalently obvious `visual' proof of the non-local character of the DDI is still missing. Such a non-ambiguous qualitative evidence of the non-local character of the dipolar interaction could be provided in principle by the observation of novel quantum phases (supersolid, checkerboard) in optical lattices~\cite{goral2002}. However, the unambiguous detection of such phases is far from trivial, as is the preparation of the ground state of the system due to a large number of metastable states~\cite{menotti2007}.

In this Letter, we investigate a minimal system, namely a mesoscopic sample of dipolar bosons in a triple-well potential, which minimizes these restrictions, while still presenting clear visual
non-local features (see ``phase'' B below). Non-dipolar BECs in double-well potentials have allowed for the observation of Josephson oscillations and non-linear self trapping~\cite{albiez2005}, showing clearly that `slicing' a BEC dramatically enhances the effects of interactions. The two-well Josephson physics is affected quantitatively (although not qualitatively) by the DDI~\cite{xiong09,pra80} (the DDI may induce however significant inter-site effects in coupled 1D and 2D bilayer systems~\cite{Arguelles2007,Wang2007,Trefzger2009}). On the contrary, as we show below, the DDI does introduce qualitatively novel physics in the Josephson-like dynamics in three-well systems. We discuss how the DDI leads to various possible ground states, which may visually reveal the non-locality of the DDI. In addition, we show how this non-locality leads to a peculiar quantum dynamics characterized by striking new phenomena, including the \emph{dynamical formation} of mesoscopic quantum superpositions (MQS). MQSs produced in cavity QED or with trapped ions~\cite{haroche2006} require complex manipulations, whereas in the present system, they arise naturally, similar to the MQSs obtained in BECs with attractive interactions in double wells~\cite{ml,ho} or lattices~\cite{mqs}. We then comment on the design of four-site Heisenberg-limited atom interferometers using the dynamical creation of MQS, and finally discuss possible experimental scenarios.

\begin{figure}[b]
\centerline{\includegraphics[width=70mm]{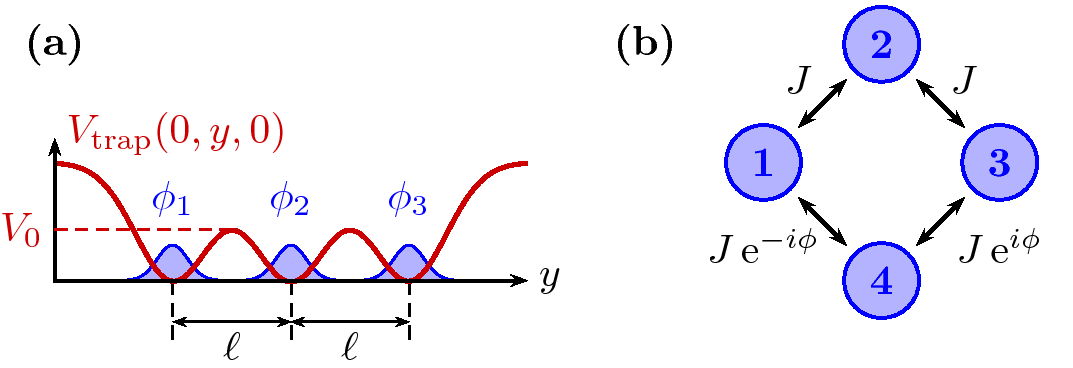}}
\caption{(color online) (a) Schematic view of the three-well system; (b) MQS interferometer with four wells (see text).}
\label{fig1}
\end{figure}

We consider $N$ dipolar bosons in a three-well potential $V_{{\rm trap}}(\bs r)$ (Fig.~\ref{fig1}a). The wells are aligned along the $y$-axis, separated by a distance $\ell$ and an energy barrier $V_0$. The bosons are polarized by a sufficiently large external field, with a dipole moment $\bs d$ along a given direction. The lattice potential is strong enough compared to other  energies (in particular the interaction energies) such that the on-site wavefunctions $\phi_{i=1,2,3}(\bs r)$ are fixed, being independent of the number of atoms per site. For a large-enough $V_0$  we may assume $\phi_i(\bs r)=\phi(\bs r-\bs r_i)$, where $\bs r_i$ is the center of site $i$. In addition we may assume $\phi$ to be a Gaussian with widths $\sigma_{x,y,z}$. We limit to the case where $\sigma_y$ is small enough with respect to $\ell$ so that the sites are well defined. Re-expressing the bosonic field operator as $\hat\psi(\bs r)=\sum_{i=1}^3 \phi_i(\bs r) \hat a_i$, we may write the Hamiltonian as:
\begin{eqnarray}
\hat H&=& -J \left [\hat a_2^\dag \left (\hat a_1+\hat a_3 \right )+ {\rm h.c.}\right ] +\frac{U_0}{2} \sum_{i=1}^{3} \hat n_i(\hat n_i-1) \nonumber \\
&+& U_1 \left [ \hat n_1\hat n_2+\hat n_2\hat n_3+\frac{1}{\alpha} \hat n_1\hat n_3 \right ],
\label{eq:H1}
\end{eqnarray}
where $J=-\int  {\rm d}{\bs r} \phi_1(\bs r)\left [-\hbar^2\nabla^2/2m+V_{\rm trap}(\bs r)\right ]\phi_2(\bs r)$ is the hopping rate, $U_0=g\int|\phi_1|^4\,{\rm d}{\bs r}+ \int|\phi_1({\bs r})|^2|\phi_1({\bs r}')|^2U_{\rm dd}({\bs r}-{\bs r}')\,{\rm d}{\bs r}\,{\rm d}{\bs r}'$ characterizes the on-site interactions, $U_1=\int|\phi_1({\bs r})|^2|\phi_2({\bs r}')|^2U_{\rm dd}({\bs r}-{\bs r}')\,{\rm d}{\bs r}\,{\rm d}{\bs r}'$ is the coupling constant for nearest-neighbor DDI, and  $\hat n_j=\hat a_j^\dag \hat a_j$. In the previous expressions $g=4\pi\hbar^2a/m$ is the coupling constant for the short-range interactions, with $a$ the $s$-wave scattering length. The DDI is given by $U_{\rm dd}({\bs r})=d^2(1-3\cos^2\theta)/r^3$, where $\theta$ is the angle between ${\bs r}$ and ${\bs d}$, $d^2\equiv\mu_0\mu^2/(4\pi)$ for magnetic dipoles ($\mu$ is the magnetic dipole moment) or $d^2\equiv {\bar d}^2/4\pi\epsilon_0$ for electric dipoles ($\bar d$ is the electric dipole moment). The parameter $\alpha$ in Eq.~\eqref{eq:H1} depends on the geometry of $V_{\rm trap}({\bs r})$ ($\alpha=8$ if the wavefunctions are well localized in all directions compared to $\ell$, and decreases towards $\alpha=4$ when $\sigma_x/\ell\to\infty$~\cite{alpha}). In the following we focus on the localized case, i.e. $\alpha=8$, but all results remain valid for $4\le \alpha\le 8$. Finally, note that $U_0$ results from short-range interactions and DDI, and that the ratio between $U_0$ and $U_1$ may be easily manipulated by means of Feshbach resonances, by modifying the dipole orientation ${\bs d}$, and by changing $\ell$~\cite{alpha}.

Since $\sum_{i}\hat n_i=N$ is conserved by~(\ref{eq:H1}), we may re-write $\hat H$ (up to a global energy $U_0 N(N-1)/2$) as an effective Hamiltonian without on-site interactions:
\begin{eqnarray}
\hat H&=& -J \left [\hat a_2^\dag \left (\hat a_1+\hat a_3 \right )+ {\rm h.c.}\right ] +(U_1-U_0)\hat n_2\left [ \hat n_1+ \hat n_3 \right ] \nonumber \\
&+&\left ( \frac{U_1}{8}-U_0\right ) \hat n_1\hat n_3.
\label{eq:Heff}
\end{eqnarray}

The gross structure of the ground-state diagram is understood from the $J=0$ case, where the Fock states $|n_1,n_3\rangle$ are eigenstates of $\hat H$, with energy $E(n_1,n_3)$ (since $N$ is conserved, the Fock states are defined by $n_{1,3}$). The minimization of $E$  provides four classical ``phases''. For $U_0>0$ and $U_1\le 8U_0/15$, and $U_0<0$ and $U_1<-8|U_0|$, phase (A) occurs, with $n_1=n_3=\lfloor{\bar n}/2\rfloor$ with
${\bar n}\equiv  16N(U_0-U_1)/(24U_0-31U_1)$ (where $\lfloor\cdot\rfloor$ denotes the integer part).
Phase (B) appears for $U_0>0$ and $8U_0/15\leq U_1 \leq 8U_0$, being characterized by $n_1=n_3=N/2$. For $U_0>0$ and $U_1>8U_0$, and $U_0<0$ and $U_1>-|U_0|$ phase (C) occurs, with $n_2=N$ (actually states with $n_{i}=N$ are degenerated, but the degeneracy is broken by tunneling which favors $n_2=N$). Finally, phase (D) occurs for $U_0<0$ and $8 U_0 <U_1< U_0 $, being characterized by a broken symmetry, with two degenerated states with $n_1=\lfloor{\bar n}\rfloor$, $n_3=0$ and vice-versa.

Fig.~\ref{fig:2}(a) shows $\langle \hat n \rangle /N$, with $\hat n=\hat n_1+\hat n_3$ for $N=18$. We can see that phases (A)--(D) describe well the gross structure of the ground-state diagram (a similar graph shows, as expected, that the (D) phase shows large fluctuations $\Delta\hat\delta$ in $\hat\delta=\hat n_1-\hat n_3$). However, tunneling is relevant at low $|U_0|$ and $|U_1|$ and at the phase boundaries. In general, the system is in a quantum superposition of different Fock states $|\psi\rangle=\sum_{n_1=0}^N\sum_{n_3=0}^{N-n_1}C(n_1,n_3)|n_1,n_3\rangle$. Fig.~\ref{fig:2}(b) depicts $\Delta\hat\delta$ in the region $U_{0,1}>0$. As expected at small $|U_{0,1}|/J$ tunneling dominates and the product state $(a_1^\dag/\sqrt{2}+a_2^\dag/2+a_3^\dag/\sqrt{2})^N |{\rm  vac}\rangle$ is retrieved ($|{\rm vac}\rangle$ is the vacuum state). This state transforms into phase (A), which for growing $U_0$ becomes the Fock state $|N/3,N/3\rangle$. Phase (C) remains the Fock state $|0,0\rangle$ ($n_2=N$), and the border (B)--(C) is characterized by a first-order ``phase transition''~\cite{transition}, at which $n_2$ abruptly jumps from $0$ to $N$. Fig.~\ref{fig:2}(c) represents schematically phases (A) to (D).

\begin{figure}[b]
\centering
\includegraphics[width=7cm,angle=0]{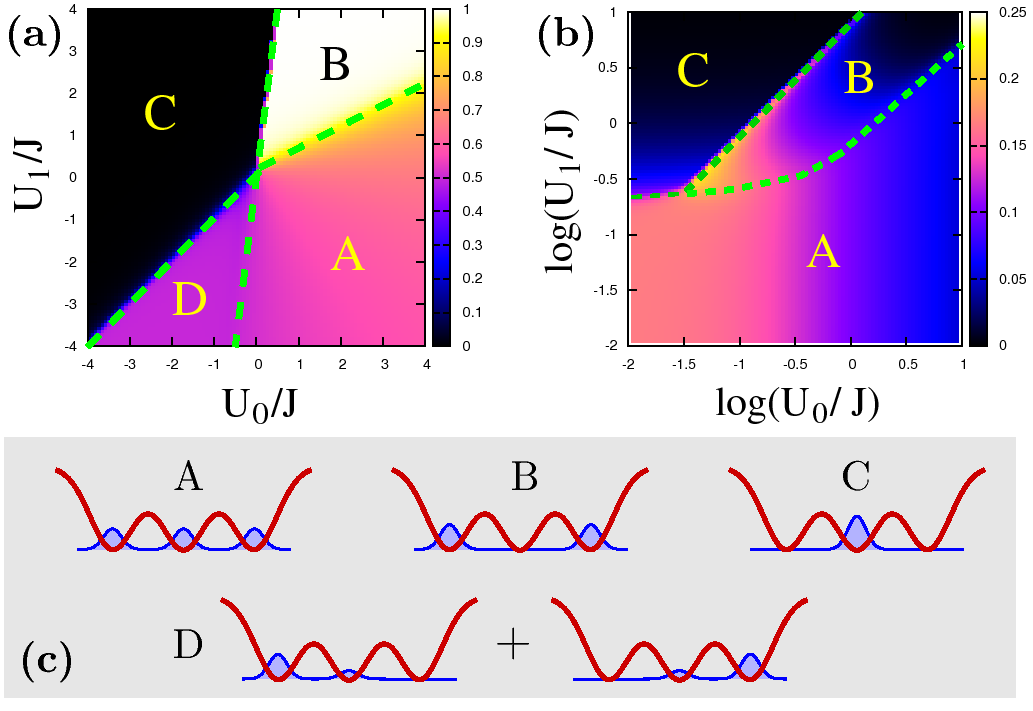}
\caption{(color online) (a) $\langle \hat n \rangle /N$ as a function of $U_{0,1}$ for $N=18$; (b) $\Delta\hat\delta/N$ in logarithmic scale for $U_{0,1}$. The dashed lines show the boundaries between the classical phases (A)--(D), that are shown schematically in (c).}
\vspace*{-0.5cm}
\label{fig:2}
\end{figure}

Phase (B) is characterized by vanishing $\langle \hat n_2 \rangle$ and $\Delta n_2 $, and $\langle \hat n_1 \rangle = \langle \hat n_3\rangle$. It strikingly reveals the non-local character of the DDI, similarly to the biconcave BECs predicted in \cite{cookie}, but with a much higher ``contrast''. Note however that the actual ground-state may significantly depart from  $|N/2,N/2\rangle$, since $|\Delta\hat\delta|$ is significant at the (B)--(C) transition (Fig.~\ref{fig:2}(b)). At $U_1=8U_0$, the ground state is a coherent state $( a_1^\dag + a_3^\dag )^N |{\rm vac}\rangle$, i.e. coherence between the two extremal sites is preserved in spite of the absence of particles in site $2$. This coherence is understood from~(\ref{eq:Heff}), since for $U_1=8U_0$ there is no effective interaction between sites $1$ and $3$. Since $\langle \hat n_2  \rangle \ll 1$ due to the effective repulsive nearest-neighbor interactions ($U_1-U_0>0$), then sites $1$ and $3$ form an effective non-interacting two-well system coherently coupled by a second-order process through site $2$ (with effective hopping $J_{{\rm eff}}=J^2/7(N-1)U_0$). Hence the coherent region extends inside (B) for $|U_1-8U_0| \lesssim J_{{\rm eff}}$. Thus for larger $NU_0$ the coherent region shrinks (reducing to the very vicinity of $U_1=8U_0$ as seen in Fig.~\ref{fig:2}(b)).

Such a $1$--$3$ coherence has important consequences for the quantum dynamics, best illustrated by considering initially all particles at site $3$. Interestingly, $\langle \hat n_{1,3} \rangle$ show perfect Josephson-like oscillations (with frequency $2J_{{\rm eff}}/\hbar$) although for any time $\langle \hat n_2 \rangle = \Delta \hat n_2 \ll 1$. However, $J_{{\rm eff}}$  decreases with $NU_0$ and hence the observation of this effect demands a mesoscopic sample, since otherwise the dynamics may become prohibitively slow. Off the $U_1=8U_0$ boundary, inside phase (B), the residual $1$--$3$ interaction leads to a damping of the Josephson oscillations (connected to number squeezing). Eventually for $|U_1-8U_0|\gg J_{{\rm eff}}$ self-trapping in $3$ occurs.

Phase (D) is characterized by a large $\Delta\hat\delta$ and $\langle \hat n_2 \rangle \neq 0$, and two degenerated states: $n_3=0$ (i), and $n_1=0$ (ii). Strictly speaking the exact ground state is provided by a MQS of these two states, but the gap between the ground-state and the first excited one is vanishingly small ($\ll J$) even at the $U_1=U_0<0$ boundary and for $N$ as small as $18$. Experimentally, the signature of phase (D)
would thus consist in measuring large shot-to-shot fluctuations in $\hat\delta$, while never observing simultaneously atoms in both sites 1 and 3. At $U_1=U_0<0$, states (i) and (ii) become coherent superpositions of the form $(a_1^\dag+a_2^\dag)^N|{\rm vac}\rangle$ and $(a_2^\dag+a_3^\dag)^N|{\rm vac}\rangle$, respectively. These superpositions may be understood from Eq.~(\ref{eq:Heff}), which for $U_1=U_0<0$ becomes
\begin{equation}
\hat H= -J \left [ \hat a_2^\dag \left ( \hat a_1+\hat a_3\right )+ {\rm h.c.}\right ] +\frac{7|U_0|}{8}\hat n_1\hat n_3.
\label{eq:Heff-2}
\end{equation}
which describes a non-interacting two-well system if $n_1=0$ or $n_3=0$, leading to the coherent states (i) and (ii).

\begin{figure}[t]
\includegraphics[width=7cm]{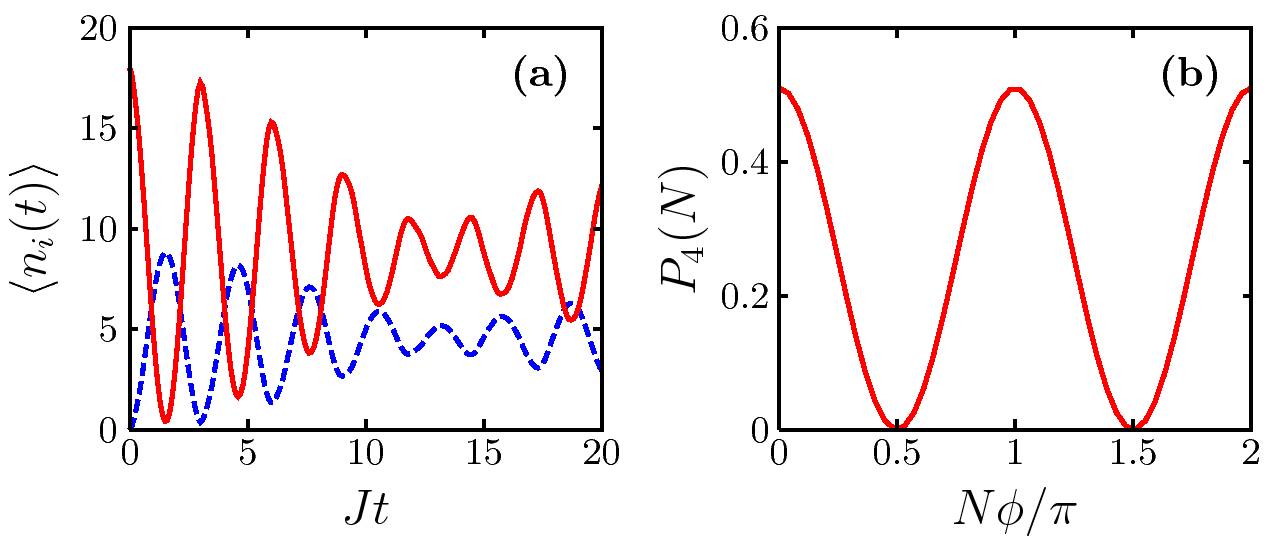}
\caption{(color online) (a) $\langle \hat n_{1,3}(t)\rangle$ (dashed), $\langle \hat n_{2}(t)\rangle$ (solid), for $U_0=U_1=-100J$ and $N=18$;
(b) Probability $P_4(N)$ as a function of $\phi$ for the interferometric $4$-site arrangement (see text) with $N=14$, $U_0=U_1=-100$ and $Jt=2.7$.}
\label{fig:3p}
\end{figure}

Hamiltonian~(\ref{eq:Heff-2}) leads to an intriguing quantum dynamics characterized by the creation of MQSs. From an initial Fock state $|0,0\rangle$ ($n_2=N$), if a particle tunnels into site $1$ (state $|1,0\rangle$), a subsequent tunneling from $2$ to $3$ (state $|1,1\rangle$) is produced with a bosonic-enhanced hopping rate $J\sqrt{N-1}$. However, the state $|1,1\rangle$ has an interaction energy $7|U_0|/8$. Hence if $J\ll 7|U_0|/8\sqrt{N-1}$ then the tunneling from $2$ to $3$ remains precluded. On the contrary the hopping into $1$ presents no energy penalty. As a result, if the first  particle tunnels into $1$, then a coherent $1-2$ superposition is established. Of course if the first particle tunnels into $3$, then a $2-3$ superposition occurs. Since the initial process is coherently produced in both directions, then a MQS $|\Phi (t)\rangle |0\rangle + |0\rangle |\Phi (t)\rangle $ is formed, where $|\Phi(t)\rangle = \sum_{n=0}^N C_n(t) |n\rangle$, with the normalization condition $2\sum_{n=1}^{N} |C_n(t)|^2+4|C_0(t)|^2=1$~\cite{alpha}. Fig.~\ref{fig:3p}a shows that $\langle \hat n_{1,3} \rangle (t)$ perform a coherent oscillation, which however damps for longer times. This damping is again a remarkable consequence of the non-local character of the DDI. Virtual hoppings of {\em a single particle} from site $2$ into site $3$ ($1$) induce a second-order correction of the energy of the states $|n,0\rangle$ ($|0,n\rangle$): $\Delta E_n=8 J^2 (N-n)/7|U_0|n $, which distorts the Josephson Hamiltonian, and leads to a significant damping after a time scale of the order of $\tau\sim 7|U_0| / 8J^2 N$ (in agreement with our numerics)~\cite{alpha}. At longer times, chaotic dynamics may even occur~\cite{chaos}.

The three-well system hence acts  as a {\em MQS-splitter} under the mentioned conditions. We stress, however, that a MQS (although asymmetric) is still created~\cite{alpha}, even for unequal hoppings $J_{ij}$ for nearest neighbors, as long as $J_{12,23}\ll 7|U_0|/8\sqrt{N-1}$. We note also that if $U_1\neq U_0$ a MQS is created if $|U_1-U_0|\lesssim J$, but nearest-neighbor interactions enhance the damping in each MQS branch. If $|U_1-U_0|\gg J$ bosons at site $2$ remain self-trapped.

The MQS-splitter opens fascinating possibilities beyond the $3$-well system, most relevantly in the context of Heisenberg-limited atom interferometry. We illustrate this possibility by considering a  simple interferometer based on a four-well system (Fig.~\ref{fig1}b). Initially the bosons are at site $2$ (which acts as the input port). Sites $1$ and $3$ play the role of the interferometer arms,
whereas site $4$ acts as the output port, where the interferometric signal is read out. We consider hoppings $J_{21}=J_{23}=J$, but $J_{34}=Je^{i\phi}=J_{14}^\star$. We are interested in the  $\phi$-sensitivity of the population at site $4$. This arrangement is chosen for its theoretical simplicity (more general arrangements work along similar lines), although it may be implemented also in  practice by means of Raman-tunneling~\cite{raman}. Under the MQS conditions (in this case $U_1=U_0<0$ and $J\sqrt{N-1}\ll (2\sqrt{2}-1)|U_0|/2\sqrt{2}$), the system evolves into an entangled MQS formed by Fock  states such that $n_i n_j=0$ for next-nearest neighbors. It is straightforward to show that the probability to find $N$ particles at site $4$ depends explicitly on the phase $\phi$ as  $P_4(N)\sim\cos^2(N\phi)$ ($P_4(n\neq N)$ are only indirectly $\phi$-dependent due to normalization). Hence $P_4(N)$ has a modulation of period $\delta\phi=\pi/N$ (Fig.~\ref{fig:3p}b), contrary to the period $\delta\phi=\pi$ expected for independent single particles, allowing
for a Heisenberg-limited interferometric measurement of the phase $\phi$. This super-resolution is an unambiguous signature of the coherent character of the MQS thus created~\cite{noon1,noon2}. $\langle \hat n_4 \rangle$ presents a similar modulation (but with poorer contrast). Calculations with a six-site arrangement provide similar results~\cite{alpha}.

In the final part of this Letter we discuss experimental feasibility. Triple-well potentials as in Fig.~\ref{fig1} may be controllably implemented with optical potentials. By superimposing, onto a single-beam optical trap which provides the $xz$-confinement, a tightly focused beam (with a waist $\sim 1\mu$m, see e.g.~\cite{sortais2008}), one may create a tight `dimple' acting as one well. To realize a triple-well (or even more complex configurations), several possibilities exist. Using an acousto-optic modulator (AOM) with several rf frequencies~\cite{shin2004,froehlich2007}, several diffracted beams are created, whose intensity and position can be controlled independently. Another option using an AOM consists in toggling the dimple between several positions at high rate, to create almost arbitrary time-averaged potentials~\cite{henderson2009}. Such an implementation has several advantages: arbitrary, time-dependent energy offsets can be applied to the different sites; the inter-site separation $\ell$ can be changed in real time, easing the preparation of a given atom number in each well (e.g. by performing evaporative cooling with different energy offsets in each site), and the detection of the population in each well (before imaging, $V_0$ may be increased to freeze out the dynamics and then $\ell$ increased, thus relaxing constraints on the imaging resolution).

We now evaluate $J$, $U_0$ and $U_1$ for realistic experimental values. Although in our calculations we have just considered $N$ up to $36$, similar ground-states are expected for larger $N$  (but, as mentioned above, the observation of the quantum features at the (B)--(C) and (D)--(C) boundaries demands small samples). In particular, consider a triple-well potential formed by three Gaussian beams of waist $1\;\mu{\rm m}$ separated by $\ell=1.7\;\mu$m. For a barrier height $V_0/h\simeq2500$~Hz, we obtain $J/h\sim 10$~Hz, and the typical value of $NU_1/J$ is then $\sim10$ for $N=2000$ $^{52}$Cr atoms. The value of $U_0$ can be tuned, for a fixed geometry, by means of Feshbach resonances~\cite{lahaye2007}, so that one can explore e.g. the first-order (B)--(C) ``transition'' with $^{52}$Cr by varying $U_1/U_0$. However, the MQS creation demands small samples, being hence more realistic with polar molecules. For example, for KRb molecules placed at a distance $\ell=1\;\mu$m and maximally polarized ($d=0.5$~D) parallel to the joining line between the sites, $U_1/h\simeq -70$Hz. Under these conditions the MQS condition implies, for $N=36$ molecules, $J/h$ of a few Hz. Single-atom sensitivity has been achieved with fluorescence imaging~\cite{imaging}, so that the relatively small values of $N$ considered here should be detectable.

In summary, we have studied a simple system of dipolar bosons in a triple well, showing that the non-locality of the DDI leads to \emph{qualitatively} novel physics that may be explored with a high degree of control over all parameters via the trap geometry, dipole orientation, and Feshbach resonances. We have shown that the ground-state phases present abrupt crossovers induced by the non-local nature of the DDI, which may be explored with $^{52}$Cr BECs. In addition, the dynamics presents intriguing new scenarios, especially for the case of polar molecules, including the dynamical creation of MQSs, which may be employed for Heisenberg-limited interferometry.

We thank M. K. Oberthaler, H. P. B\"uchler, D. Gu\'ery-Odelin, and E. Demler for useful discussions. We acknowledge support by the DFG (QUEST and SFB/TRR 21), the ESF (EUROQUASAR) and the EU (Marie-Curie Grant MEIF-CT-2006-038959 to T. L.).

\newpage

\onecolumngrid
\noindent
\begin{center}
{\large \bf Supplementary material for \\ ``Mesoscopic ensembles of polar bosons in triple-well potentials''}
\end{center}
\vskip 1cm
\twocolumngrid

In this supplementary text, we provide additional details concerning
several points discussed in the Letter.

\section{Variation of the parameter $\alpha$ with the ratio $\ell/\sigma_x$}

\begin{figure}[b]
\includegraphics[width=8.3cm,angle=0]{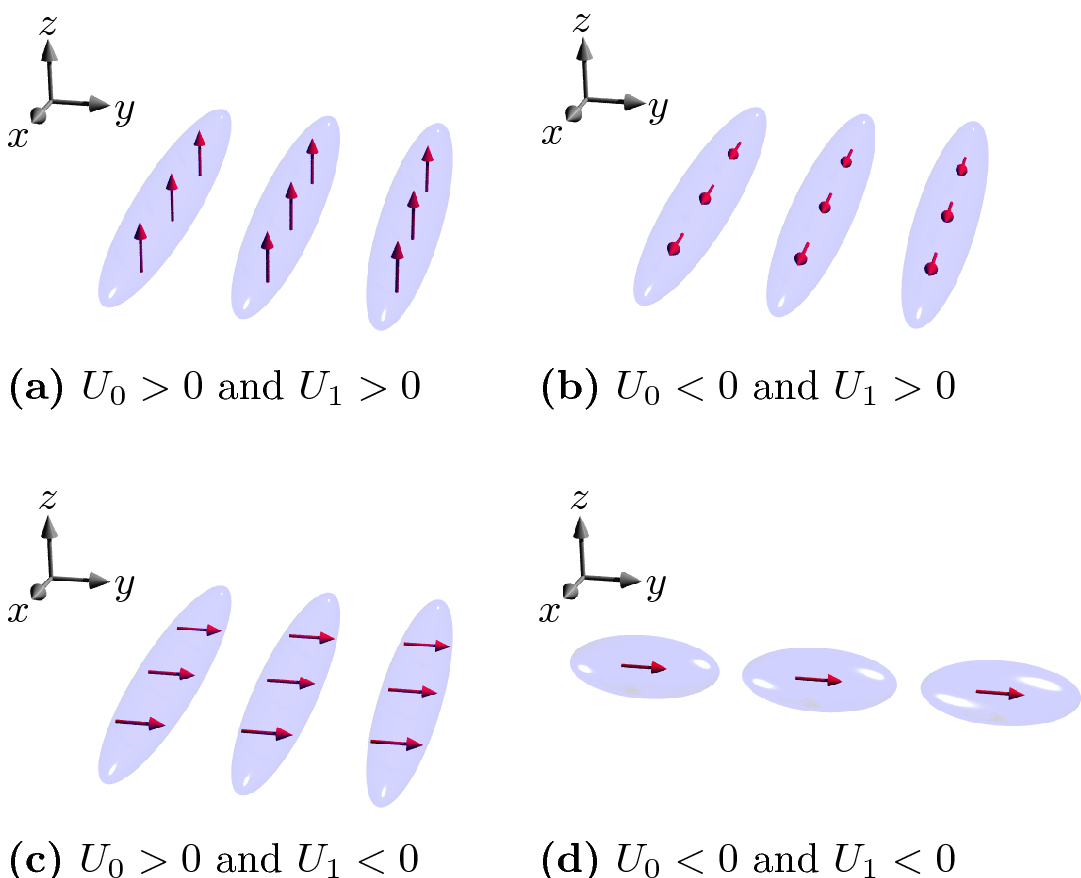}
\caption{Changing the geometry of the sites and the orientation of the dipoles allows, for three aligned wells, to choose any combination of signs for $U_0, U_1$. For cases (b) and (c), the signs given on the figure are correct only for moderate aspect ratios of the on-site wavefunctions.}
\label{fig:tune}
\end{figure}

The parameter $\alpha$ is defined as the ratio
\begin{equation}
\alpha=\frac{E(\sigma_x,\ell)}{E(\sigma_x,2\ell)}
\label{eq:def:a}
\end{equation}
where $E(\sigma_x,\ell)$ is the dipolar interaction energy of two Gaussian clouds of length $\sigma_x$ along the $x$-axis, containing $N$ particles, and separated by a distance $\ell$ along the $y$ direction. Here we consider that the dipoles are pointing towards the $z$ direction as in Fig.~\ref{fig:tune}(a) (implying $U_1>0$), but the results hold for other dipole orientations. We assume that the sizes $\sigma_{y,z}$ of the cloud along the $y$ and $z$ axes are small as compared to $\ell$. Therefore, $E$ depends only on $\ell$ and $\sigma_x$. From scale invariance, we deduce  that $\alpha$ depends only on the ratio $\ell/\sigma_x$. We can easily work out two limits: (i) if $\sigma_x\ll \ell$, the two clouds can be considered as point-like, and the dipolar energy of the two clouds thus scales as $E\sim 1/\ell^3$. In that case, one thus gets, from (\ref{eq:def:a}), $\alpha=8$; (ii) in the opposite case $\sigma_x\gg \ell$, one practically deals with two infinite chains of dipoles, and the interaction energy is then known to scale as $E\sim1/\ell^2$, yielding $\alpha=4$. In between these two cases, $\alpha$ varies monotonously with the ratio $\ell/\sigma_x$, as we shall see below.

\begin{figure}[b]
\includegraphics[width=6.0cm,angle=0]{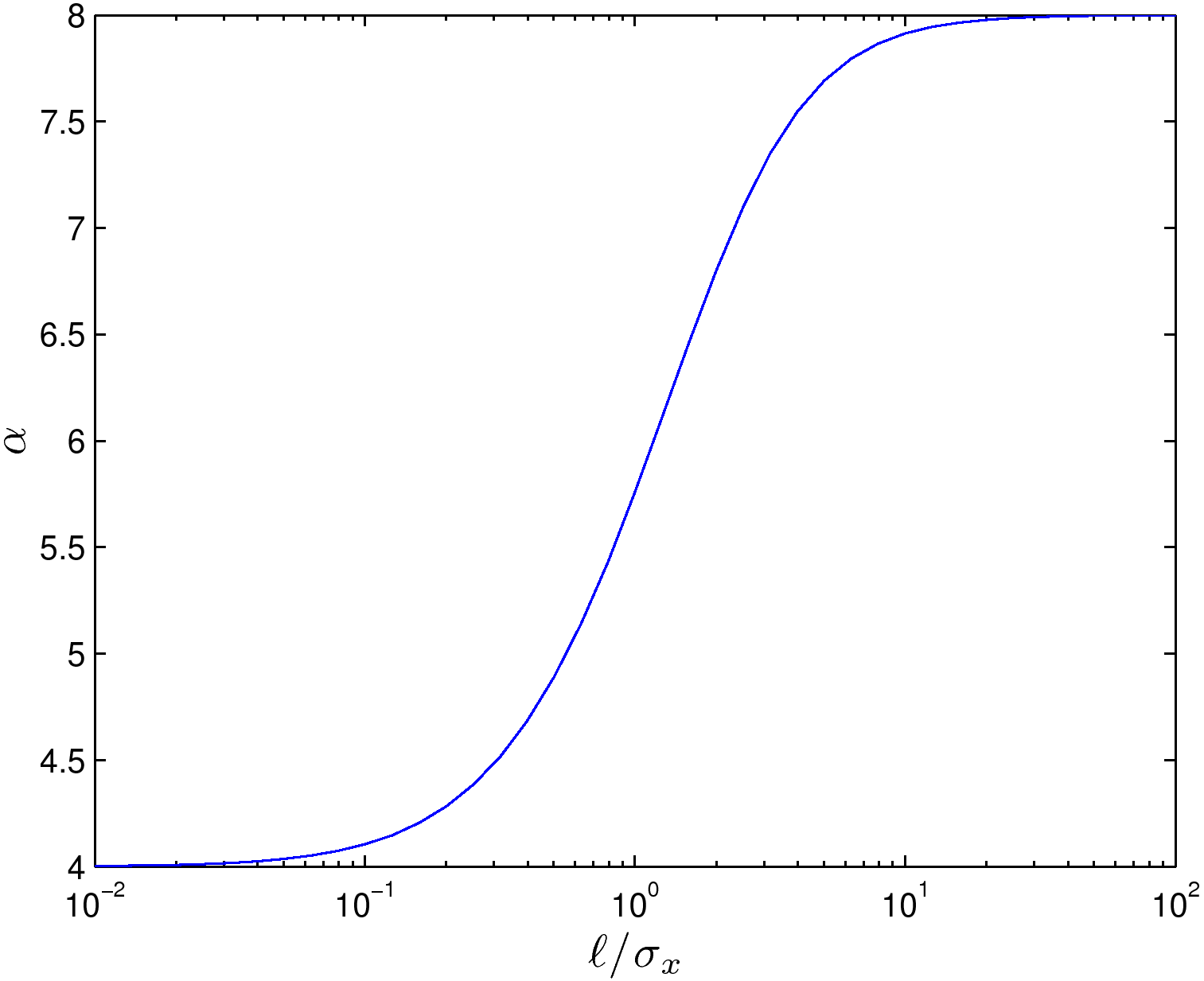}
\caption{Parameter $\alpha$ as a function of $\ell/\sigma_x$, obtained from (\ref{eq:fig}).}
\label{fig:alpha}
\end{figure}

Since we assume $\sigma_{y,z}$ small compared to $\ell$, we can approximate the Gaussian atomic densities by:
\begin{eqnarray*}
n_1(x,y,z)&=&A\exp\left(-\frac{x^2}{\sigma_x^2}\right)\,\delta(y)\,\delta(z),\\
n_2(x,y,z)&=&A\exp\left(-\frac{x^2}{\sigma_x^2}\right)\,\delta(y-\ell)\,\delta(z),
\end{eqnarray*}
where $A$ is a normalization constant. Calculating the dipolar interaction energy
$$
E=\int {\rm d}{\bs r}\,{\rm d}{\bs r}'\,n_1({\bs r})n_2({\bs r}')U_{\rm dd}({\bs r}-{\bs r}'),
$$
we get
\begin{equation}
E(\sigma_x,\ell)\propto\int{\rm d}x\,{\rm d}x'\;\frac{\exp\left[-\left(x^2+x'^2\right)/\sigma_x^2\right]}{\left[(x-x')^2+\ell^2\right]^{3/2}}.
\label{eq:fig}
\end{equation}
From this expression, it is clear that $\alpha$ depends only on the ratio $\ell/\sigma_x$, as expected from scale invariance; this dependence is shown on Fig.~\ref{fig:alpha}, where the curve was obtained by a numerical integration of~(\ref{eq:fig}).

\section{Tuning of the interaction parameters $U_0$ and $U_1$}

In this section we elaborate on how the interaction parameters can be modified such that
the sign of $U_{0,1}$ can be tuned. First, let us consider the arrangement of three aligned wells. Disregarding the contact interaction contribution to $U_0$, all sign combinations for $U_{0,1}$ can be achieved by exploiting the anisotropy of the DDI in a carefully designed geometry, as sketched in Figs.~\ref{fig:tune}.

We now turn to the the square (or even hexagonal) arrangement for the interferometric setup
(see the Letter and sections below), for which we want to have $U_0<0$ and $U_1<0$. In this case, because of geometric constraints, the sign of the DDI must be inverted. In principle, this can be achieved for magnetic dipoles by using a rotating field~\cite{tune} (although this precludes the use of magnetically induced Feshbach resonances). In the case of polar molecules (more relevant for the study of the MQS splitter as discussed in the Letter), dressing the rotational levels by AC fields, as proposed in~\cite{micheli2007}, allows for inverting the sign of the DDI; in combination with a proper design of the trapping geometry, this should allow for the realization of the proposed experiments.

\section{Dynamics of the MQS splitter}

In this section we comment on the dynamics of the MQS-splitter in detail.
Under the MQS conditions $U_1=U_0<0$ and $J\sqrt{N-1}\ll 7|U_0|/8$,
only states of the form (using the notation of the Letter) $|n,0\rangle$ or
$|0,n\rangle$ are possible. Under these conditions the system (originally prepared
in $|0,0\rangle$) evolves dynamically into a MQS state of the
general form:
\begin{equation}
|\psi(t)\rangle =  |\Phi\rangle |0\rangle + |0\rangle  |\Phi\rangle,
\end{equation}
with $|\Phi\rangle \equiv \sum_{n=0}^N C_n(t) |N-n\rangle$, where the $C_n(t)$ coefficients
fulfill the normalization condition $2\sum_{n=0,N-1} |C_n(t)|^2+4|C_N(t)|^2=1$.

\begin{figure}[t]
\centering
\includegraphics*[width=6cm]{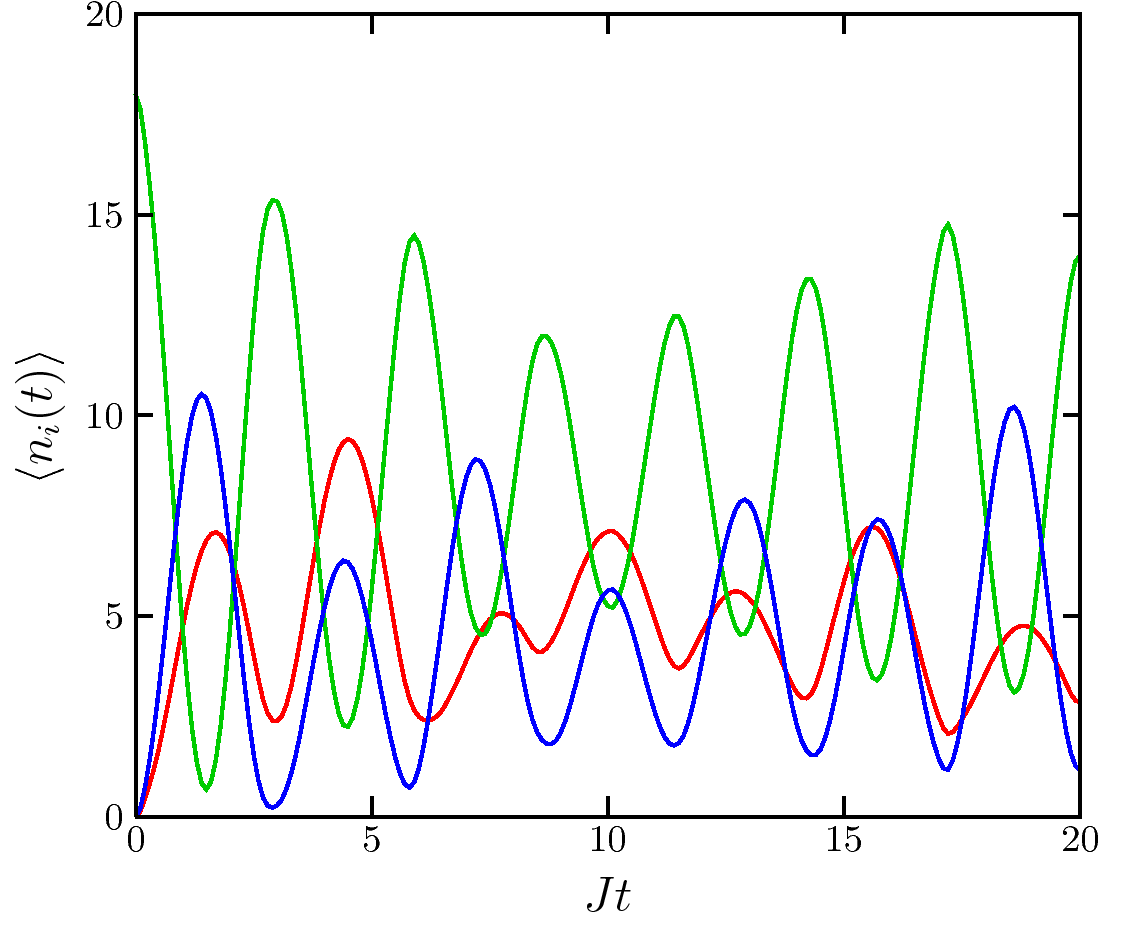}
\caption{Averages $\langle \hat n_1\rangle$ (red), $\langle \hat n_2 \rangle$ (green) and
$\langle\hat n_3\rangle$ (blue) for $U_0=U_1=-100J$ and $J_{12}=0.9J$, $J_{23}=1.1J$.}
\label{fig:J12J23}
\end{figure}

Note that a slightly different tunneling $J_{12}=J+\delta J$, $J_{23}=J-\delta J$ such that
both $J_{12,23}\sqrt{N-1}\ll 7|U_0|/8$ will lead to a similar state, although now
slightly asymmetric:
\begin{equation}
|\psi(t)\rangle =  |\Phi\rangle |0\rangle + |0\rangle  |\Phi'\rangle.
\end{equation}
In Fig.~\ref{fig:J12J23} we consider the case $J_{12}=0.9J$ and $J_{23}=1.1J$, $U_0=U_1=-100J$. One may see that the dynamics is, as expected, different for sites $1$ and $3$ (this should
be compared with Fig. 3(a) in the Letter). However, the state is still of the MQS form, since $\langle \hat n_1 \hat n_3 \rangle$ remains negligible ($<3\times 10^{-3}$) at any time.

Let us go back to the symmetric case $J_{12}=J_{23}=J$. Fig. 3b in the Letter shows the evolution
of the populations for $U_0=-100 J=U_1$, and $N=18$ particles. We have evaluated that the MQS condition
is actually fulfilled, i.e. $\langle \hat n_1 \hat n_3\rangle$ remains negligible at all times.
The population in $1$ follows a coherent Josephson oscillation, but it damps after several oscillations
(for $U_0=-10 J$, only one oscillation remains). The reason behind this damping is interesting, since
it constitutes itself a quite remarkable non-local effect introduced by the dipole-dipole interactions.

The main reason behind this damping is the existence of second-order processes, coming from virtual transitions of the form $|0,n\rangle \rightarrow |1,n\rangle \rightarrow |0,n\rangle $, i.e.
a single particle tunnels to the left from site $2$ and comes back. These virtual explorations of the left side
(by a single particle!) result in an energy shift for the state $|0,n\rangle$, which acquires
a second order energy shift $ 8 J^2 (N-n)/7|U_0|n $. This means that there is a
coupling between different many-body states which have different energies. The physics of the $2$-$3$ system is then given by the effective Hamiltonian (here $|n\rangle$ means only the site $3$)
\begin{eqnarray}
\hat H_{\rm eff} &\simeq& -J\sum_{n=1}^N \sqrt{n} \left [
|n-1\rangle\langle n|+|n\rangle\langle n-1|
\right ] \nonumber \\
&+&\sum_{n=1}^N \frac{8 J^2 (N-n)}{7|U_0|n} |n\rangle\langle n|.
\end{eqnarray}
Note that without the extra second-order correction we have the usual Josephson coupling (no interactions).
But with the second-order term the problem is now truly many-body, and we cannot write the Hamiltonian
any more as a Josephson-like Hamiltonian. This second-order shift is the responsible of
the observed damping. This may be easily seen by having a look to the typical
energy scale of the perturbation $ \sim 8J^2 N/7|U_0|\simeq 0.18 J$ (for the example of Fig. 3b of the Letter), which
leads to a damping time scale of the order of $6/J$, which is what one observes in the numerics.

Summarizing, the system behaves indeed as a MQS-splitter.
Second-order processes (due to single-particle tunnelings to the left for right moving states,
or viceversa for left-moving states) lead to a damping of the Josephson oscillations, and hence induce a
complex and rich dynamics for the MQSs. This has important practical consequences, e.g. there are optimum times
for the interferometric measurements (see below).

\section{Interferometric arrangement using a four-site square system}

In this section we would like to discuss the use of the MQS-splitter idea in the context of
Heisenberg-limited interferometry.

We consider a $4$-site arrangement as in Fig.~1(b) of the Letter. There is tunneling between nearest neighbors with hopping constant $J$. However, we shall introduce an additional phase $\phi$, such that the tunneling $J_{3\rightarrow 4}=e^{i\phi}J$ and $J_{1\rightarrow 4}=e^{-i\phi}J$. We are interested in the sensitivity of the system to this phase. We will consider that the initial population is all in site $2$. We shall now evaluate the $\phi$-sensitivity of the population at site $4$.

The Hamiltonian of the system is of the form: $\hat H=\hat H_T+\hat H_{I}$, where the hopping part
is given by
\begin{eqnarray}
\hat H_T&=& -J \left [ \hat a_2^\dag (\hat a_1+\hat a_3)+ {\rm h.c.}\right ] \nonumber \\
&-& J \left [\hat a_4^\dag (e^{i\phi} \hat a_3  +e^{-i\phi} \hat a_1) + {\rm h.c.} \right ],
\end{eqnarray}
and the interaction part is
\begin{eqnarray}
\hat H_I &=& \frac{U_0}{2} \sum_{i=1}^{4} \hat n_i(\hat n_i-1) \nonumber \\
&+&\!\!\!\! U_1 \left [ (\hat n_1+ \hat n_3)(\hat n_2+\hat n_4)\! + \!\frac{1}{2\sqrt{2}} (\hat n_1\hat n_3+\hat n_2\hat n_4) \right ].
\end{eqnarray}
As for the triangle, we may eliminate a constant and get the effective interaction part:
\begin{eqnarray}
\hat H_I&=& (U_1-U_0)(\hat n_1+\hat n_3)(\hat n_2+\hat n_4) \nonumber \\
&+&\left (\frac{U_1}{2\sqrt{2}}-U_0\right ) (\hat n_1\hat n_3+\hat n_2\hat n_4).
\label{eq:Heff}
\end{eqnarray}
Note that now the sites are at the vertices of a square, and hence the next-nearest-neighbor interaction
is $U_1/2\sqrt{2}$ (as for the $3$-site case we assume here point-like sites, see the discussion of the
first section above).

We may again find the MQS-splitter condition $U_1=U_0<0$. In that case
\begin{equation}
H_I=\left ( \frac{2\sqrt{2}-1}{2\sqrt{2}} \right )|U_0| (\hat n_1\hat n_3+\hat n_2\hat n_4).
\end{equation}
We have again a very similar MQS-splitter condition as that in the Letter:
\begin{equation}
J\sqrt{N-1}\ll \left ( \frac{2\sqrt{2}-1}{2\sqrt{2}} \right )|U_0|.
\end{equation}
When this is fulfilled, then starting from all the population in site $2$, if $1$ is populated $3$ is not, and viceversa.
Moreover, due to the (effective) repulsion between sites $2$ and $4$, $4$ may be only populated once $2$ is empty.

This means that the state of the system is of the form (with the Fock-state notation $|n_1,n_2,n_3,n_4\rangle$):
\begin{widetext}
\begin{eqnarray}
|\psi(t)\rangle &=& c(t)\sum_n D_n(t) (|N-n,n,0,0\rangle + |0,n,N-n,0\rangle) \nonumber \\
&+& s(t)\sum_n C_n(\phi,t)(e^{in\phi}|0,0,N-n,n\rangle + e^{-in\phi}|N-n,0,0,n\rangle),
\end{eqnarray}
\end{widetext}
with the normalizations:
\begin{eqnarray}
|c(t)|^2+|s(t)|^2&=&1, \\
2\sum_{n\neq N} |D_n(t)|^2+4|D_N(t)|^2&=&1,\\
2\sum_{n\neq N} |C_n(\phi,t)|^2+4|C_N(\phi,t)|^2 \cos^2 N\phi&=&1.
\end{eqnarray}

In principle the $\phi$ dependence appears explicitly only for the $n_4=N$ term,
but note that due to the normalization condition it appears also in terms with $n_4\neq N$.
The probability to find all the atoms at site $4$ is hence $P_4(N)=4|C_N(\phi,t)|^2 \cos^2 N\phi$,
and hence it has a modulation of period $\delta\phi=\pi/N$. The modulation of $P_4(N)$ has
clearly a contrast $1$ (see Fig. 3(b) of the Letter). The population in $4$ is hence
very sensitive to the phase, and, if monitored, allows for a Heisenberg-limited interferometric
measurement of the phase $\phi$. Obviously, if, instead of the above coherent superposition, one creates a statistical mixture, the probability to find all the particles at site~4 is independent of $\phi$.

\begin{figure}[t]
\centering
\includegraphics*[width=70mm]{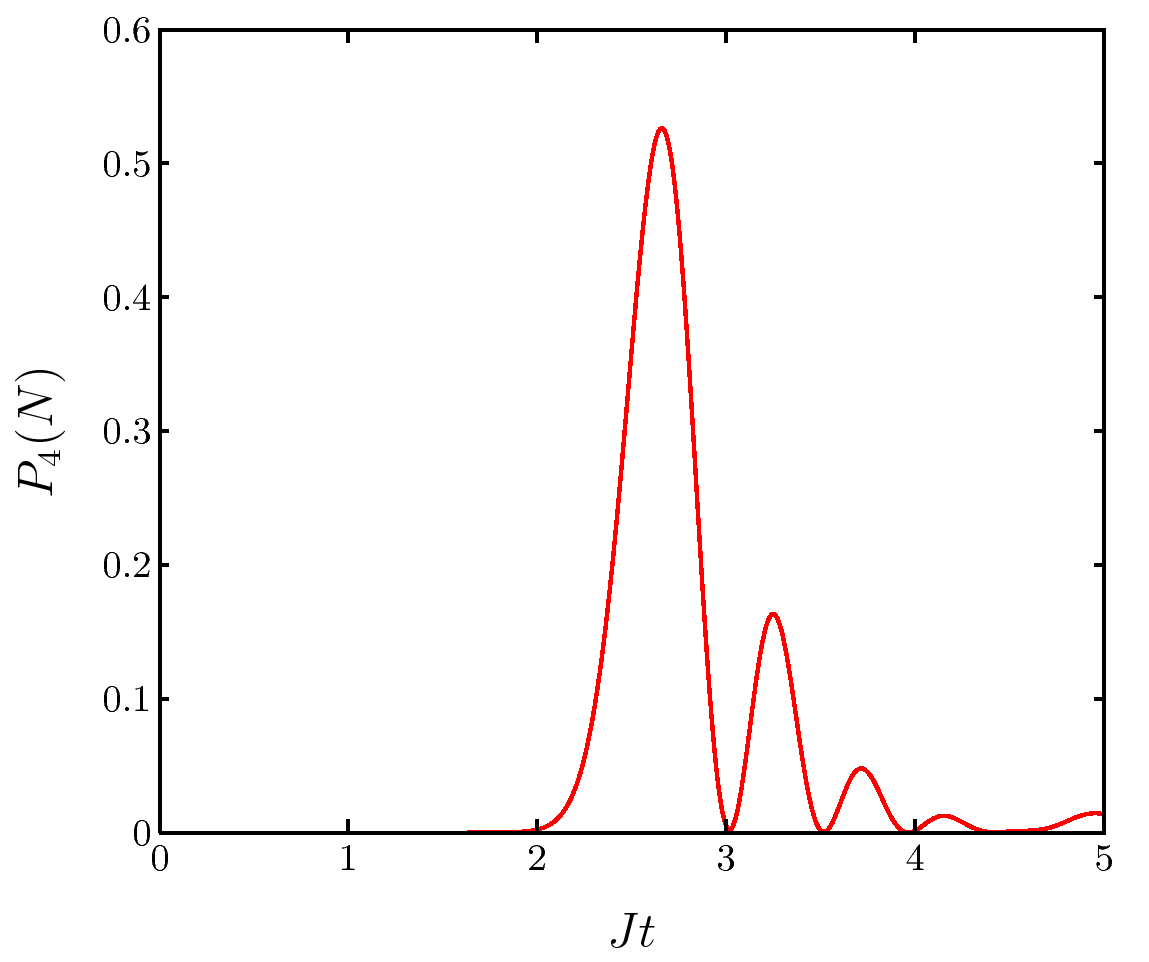}
\caption{$P_4(N)$ as a function of $Jt$ for $N=14$ and $U_0=U_1=-100J$.}
\label{fig:4sites-dynamics}
\end{figure}

In Fig. 3(b) in the Letter we have chosen the time $Jt=2.7$ for the interferometric measurement.
This is done so, since for this time the transfer into site $4$ is optimal, i.e. the probability $P_4(N)$
is maximal for $\phi=0$ (see Fig.~\ref{fig:4sites-dynamics}).

\begin{figure}[t]
 \centering
 \includegraphics[width=7cm]{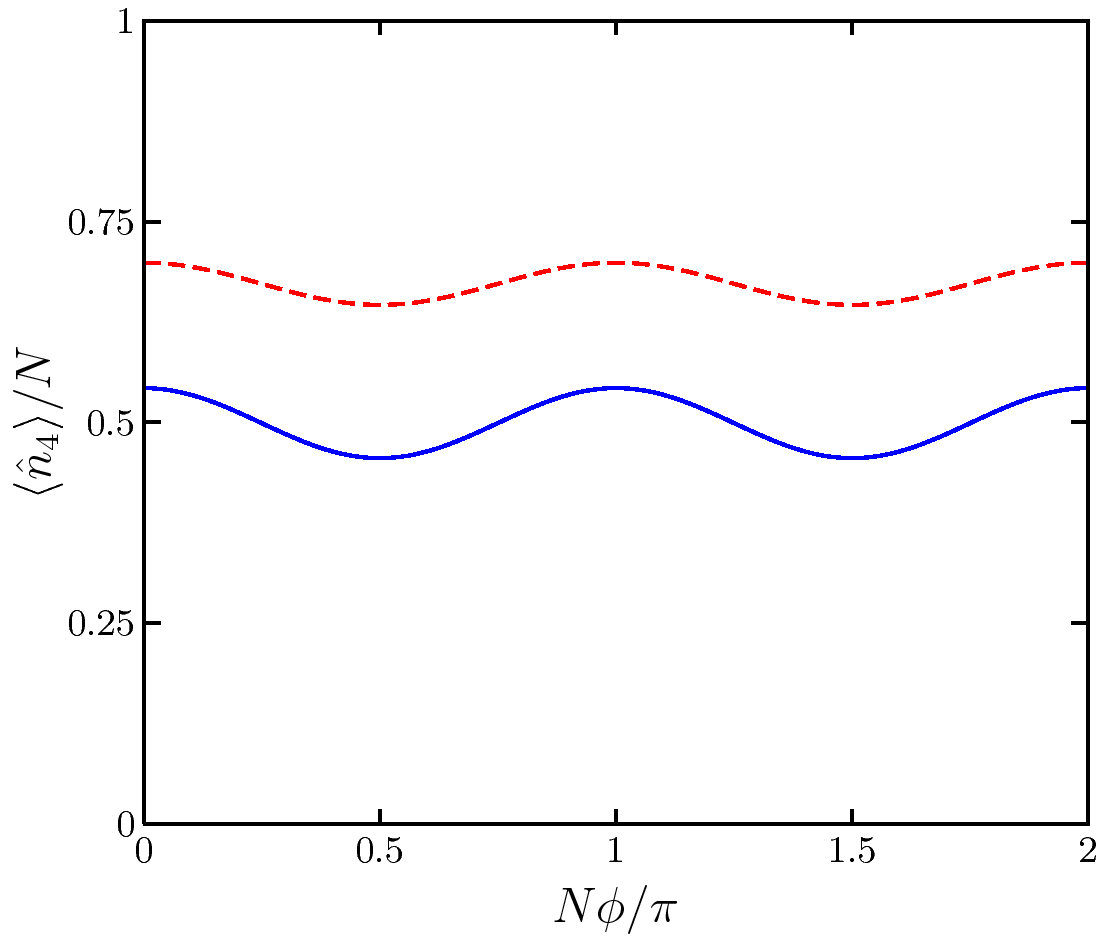}
 \caption{Average population $\langle \hat n_4 \rangle/N$ as a function of $N\phi/\pi$ for the
$4$-site system with $N=14$, $U_0=U_1=-100J$ after $Jt=2.7$ (dashed red curve), and for the
$6$-site system (solid blue curve), at time
$Jt=3.48$ and same $U_{0,1}$.}
 \label{fig:2}
 \end{figure}

One may also monitor the average population $\langle \hat n_4 \rangle$. This population is also
modulated (see Fig.~\ref{fig:2}) but the contrast is poorer. The reason may be found in the
normalization condition, since when the population $P_4(N)$ tends to zero, this increases the
relative probability to occupy $n_4\neq N$. The latter may be observed by monitoring
the population distribution $P_4(n)$. Note also that, again due to normalization, $\langle \hat n_{1,3}\rangle$
are also modulated (out of phase) with period $\delta\phi=\pi/N$.

These results must be compared with the non-interacting case $U_0=U_1=0$, where $P_4(N)$
shows a (trivial) modulation with period $\delta\phi=\pi$.

\section{Hexagonal arrangement}
\begin{figure}[h!]
\centering
\includegraphics*[width=4cm,angle=0]{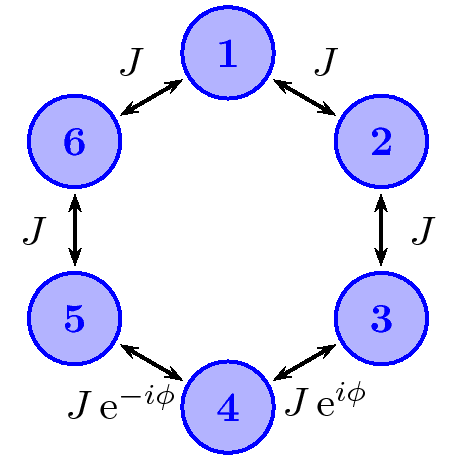}
\caption{Sketch of the $6$-site hexagonal system employed as a simple model of an atom interferometer.}
\label{fig:6sites}
\end{figure}

In this last section we briefly discuss an hexagonal interferometric arrangement, showing that the
Heisenberg-limited interferometry based in the MQS-splitter may be extended to more general setups.
We consider the arrangement shown in Fig.~\ref{fig:6sites}. The full quantum simulation of the dynamics
becomes much harder, and may be only performed for a rather small number of particles. We have performed
simulations for up to $N=6$ particles. Initially all the population is at site $1$ (input port of the interferometer).
Sites $2-3$ and $6-5$ act as the arms of the interferometer, whereas site $4$ acts as the output port, where the
interferometric measurement is performed. As for the $4$-site arrangement all hoppings are the same, except for the hoppings
$J_{34}$ and $J_{54}$ which have an extra phase. $P_4(N)$ presents a perfect modulation
with period $\delta\phi=\pi/N$, very similar as for the $4$-site system (Fig. 3(b) in the Letter).
In Fig.~\ref{fig:2} we show the comparison for $4$ and $6$ sites of $\langle \hat n_4\rangle/N$
a a function of $N\phi/\pi$.

\begin{thebibliography}{99}
\bibitem{transition}
Strictly speaking, one deals only with a crossover, since the system has a finite size.
\end{thebibliography}
\end{document}